# COMMUNICATION

## Unsymmetrical synthesis of benzimidazole-fused naphthalene imides with panchromatic absorption and redox activity

Guan-Ru Lin,[a]† Huai-Chih Chang,[a]† Yi-Chen Wu,[a] Chen-Kai Hsieh,[a] Chih-Jou Chien,[a] Guan-Lin Lu,[a] Makeshmuralikrishna Kulasekaran,[b] Milanmathew Sssuraj,[c] Tzu-Ling Ho,[a] Jatin Rawat,[d] and Hsien-Hsin Chou,*[a]

**We report a concise synthesis of unsymmetrical benzimidazole-fused naphthalene imide (BfNI) and anhydride (BfNA) derivatives featuring broad UV–Vis–NIR absorption, stable redox activity, and enhanced solubility. Incorporation of triarylamine donors induces strong intramolecular charge transfer and narrows the optical bandgap. This modular design bypasses multistep protection-deprotection and complex π-assembly, offering a versatile platform for tunable optoelectronic materials.**

Organic semiconducting materials are essential for optoelectronic devices due to their tunable π-conjugated frameworks.[1] Rylene imides[2–4]—especially naphthalene-1,4,5,8-tetracarboxylic diimides (NDIs) and perylene-3,4:9,10-tetracarboxylic diimides (PDIs)—are valued for their redox activity, strong visible absorption, and chemical robustness.[3,5–7] Their structural versatility allows for modulation of electronic properties and solubility through functionalization at the π-core (e.g., core positions in NDIs and bay/ortho positions in PDIs), as well as at the peri and imide termini. These imides and their dianhydride counterparts, e.g., naphthalene (NTCDA) and perylene (PTCDA) analogues, have been widely adoped in solar cells,[2,3,7,8] field-effect transistors,[6,9–13] sensors,[2] and photocatalysis,[14–16] and have also emerged as promising redox-active materials for energy storage.[5,17–20]

Materials with broadened absorption from the UV to near-infrared (NIR) regions are highly desirable for enhancing light-harvesting efficiency in photovoltaic devices.[21] π-Extension strategies, including peri-fused naphthalene units, donor-acceptor engineering, axial functionalization, heteroatom introduction, and π-core annulation, have been widely explored[3,4,11,14,22,23] to narrow bandgaps and red-shift absorption.[24,25] However, achieving truly panchromatic absorption in rylene dyes remains challenging. While these strategies effectively extend absorption into the NIR region, they often present a trade-off by introducing spectral "vacancies" in the blue-green region.[3,4,26–29] Alternatively, various low-bandgap materials, exemplified by porphyrins and BODIPYs,[21,30] can extend spectral coverage but typically at the cost of synthetic complexity and reduced functional flexibility.

Although rare, true panchromatic absorption has been achieved in select systems. For instance, molecular hybrids connecting perylene monoimide (PMI) with porphyrin via ethynylene linker offer uniform 400–700 nm coverage with high extinction coefficients (~43,000 $M^{-1}cm^{-1}$).[31] DPP-*N*-annulated PDI/BTXI conjugates (where DPP and BTXI denote diketopyrrolepyrrole and donor-like benzothioxanthene-3,4-dicarboximide, respectively) exhibit broad absorption from 300 to 750 nm, although further structural modification is constrained by synthetic limitations.[32] Despite successful absorption tuning, their rigid and planar backbones often lead to aggregation and poor solubility due to π–π stacking, limiting functionalization and processability.[33] Benzimidazole-fused rylene imides, such as perinone and PTCDI, also offer strong light harvesting and redox stability.[9,34–36] In contrast, prior studies mainly focus on symmetric structures, limiting tunability. Unsymmetric π-fusion at mono-shoulder or axial positions remains underexplored due to synthetic challenges.[35,37–39]

Herein, we present a new approach for unsymmetric rylene imides that addresses key limitations while preserving synthetic simplicity, true panchromaticity, and redox stability. benzimidazole-fused naphthalene imide (BfNI) and anhydride (BfNA) derivatives were synthesized via unsymmetrical π-fusion and donor-acceptor design, achieving enhanced solubility. Broadband absorption extending into the NIR region were also

a. *Department of Applied Chemistry, Providence University, Taichung 433303, Taiwan.*
b. *Department of Chemistry, SRM Institute of Science and Technology, Kattankulathur, Tamil Nadu 603203, India.*
c. *Department of Chemistry, Karpagam Academy of Higher Education, Coimbatore, Tamil Nadu 641021, India.*
d. *Department of Chemical Engineering, Institute of Chemical Technology Mumbai IndianOil Odisha Campus, Bhubaneswar, Odisha 751013, India.*
† These authors contributed equally.

achieved via breaking the molecular symmetry. This approach offers a streamlined alternative to conventional symmetric π-extension strategies at the core, shoulder, or axial positions of naphthalene imides.[2,11,13,16,29,40]

Scheme 1 outlines the synthesis of benzimidazole-fused naphthalene imide (BfNI) derivatives **C3-HC** and **C3-GR** via an unsymmetric route. Starting from commercially available NTCDA, mono-imidation with cetylamine in DMF at 140 °C afforded the naphthalene monoimide monoanhydride (**C1**) in good yield. The long alkyl chain imparts minimal yet sufficient solubility to suppress double imidation and allows isolation without protecting groups. It also enhances compatibility with common organic solvents such as $CH_2Cl_2$, THF, and ethyl acetate. This streamlined method bypasses conventional multistep protection/deprotection strategies for rylene core functionalization.[41] Subsequent condensation of **C1** with 4,5-dibromobenzene-1,2-diamine (**B3**, see Scheme S1 in Supporting Information) in hot DMF afforded **C2**, the primary benzimidazole-fused naphthalene imide (BfNI) moiety bearing two bromine substituents. Cross-coupling of **C2** with *N,N*-bis(4-alkoxyphenyl)aniline boronates (**A4** or **A4'**) via Pd-catalyzed Suzuki–Miyaura conditions furnished the BfNI derivatives **C3-GR** and **C3-HC**, respectively (Scheme 1, Method A). Alternatively, these π-extended targets could be accessed directly from **C1** via one-pot condensation with pre-functionalized *o*-diaminoarene (**B6/B6'**), synthesized from 5,6-dibromobenzo[2,1,3]thiadiazole (**B4**) and the same boronates, followed by zinc-mediated benzo[2,1,3]thiadiazole (BTD) reduction (Method B, see also Supporting Information). Although Method B affords a slightly higher overall yield, it requires handling of the unstable bis(triarylamine)-substituted intermediate **B6/B6'**, making the purification process more laborious and costly than that of

Method A. Furthermore, the lateral imide group on **C3-HC** was converted to the corresponding anhydride derivative, **C4-HC**, under basic conditions. Alternatively, direct condensation of NTCDA with **B6** also yielded **C4-HC**, as confirmed by TLC and crude NMR analysis. However, this route suffers from low yield and purification difficulties due to the formation of complex side products.

The [1]H NMR spectra of both BfNI and BfNA derivatives display clear splitting of the four aromatic protons on the naphthalene core, appearing as two doublets of doublets near 9.0 ppm, indicative of unsymmetric substitution at shoulder or axial positions. In the [13]C NMR spectra, the rigid π-conjugated backbone and multiple tertiary carbons lead to extended spin-lattice relaxation times ($T_1 > 3.5$ s), reflecting reduced local magnetic interactions and a highly delocalized electronic structure.

The BfNI derivatives exhibit pronounced intramolecular charge transfer (ICT) bands and strong molar absorptivity across the visible and near-infrared regions. As shown in Figure 1(a), the dibromo precursor **C2** absorbs maximally at 434 nm ($\varepsilon$ = 225,000 $M^{-1}$ $cm^{-1}$) in $CH_2Cl_2$, already surpassing the dicyano analogue[39] ($\lambda_{max}$ = 424 nm, $\varepsilon$ = 25,000 $M^{-1}$ $cm^{-1}$) and reference V-shaped push-pull type chromophore **B5** ($\lambda_{max}$ = 430 nm, $\varepsilon$ = 77,000 $M^{-1}$ $cm^{-1}$), which incorporates a BTD bridge. Incorporation of triarylamine groups in **C3-HC** leads to dual-band absorption: a sharp ICT band at 436 nm ($\varepsilon$ = 23,000 $M^{-1}$ $cm^{-1}$) and a broad band extending to 565 nm ($\varepsilon$ = 6,000 $M^{-1}$ $cm^{-1}$), with an onset reaching approximately 750 nm. The same behavior was observed for **C3-GR**. Notably, the anhydride derivative **C4-HC**, bearing closely related benzimidazole-fused naphthalene backbone (BfNA), also shows a red-shifted shoulder at 434 nm ($\varepsilon$ = 18,000 $M^{-1}$ $cm^{-1}$) and a flattened band near 563 nm ($\varepsilon$ = 6,000 $M^{-1}$ $cm^{-1}$), with an absorption edge extending to 750 nm. Additionally, BfNI/BfNA absorption matches indoor LED emission, suggesting potential for ambient-light use. Our molecular design significantly outperforms the $\lambda_{max}$ values for the naphthalene imides with symmetrical substituted amines (620 nm, onset at ca. 675 nm),[42] fused indoles (580-622 nm, onset at 640-675 nm),[43] and fused thiophenes (550 nm, onset at ca. 575 nm),[44] all of which lack significant absorption in the blue-green region. These results

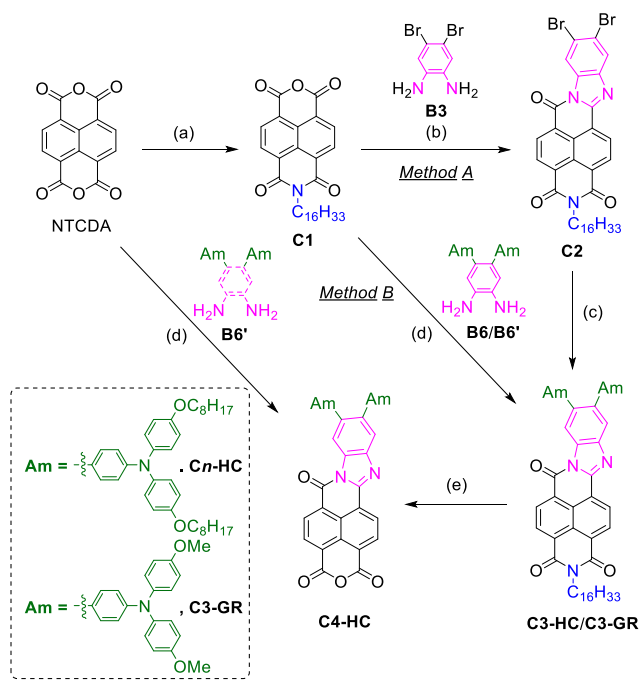

**Scheme 2** Synthetic routes for BfNI and BfNA derivatives. Conditions: (a) $C_{16}H_{33}NH_2$, DMF, 140 °C. (b) DMF, 140 °C. (c) Am-Bpin (**A4** or **A4'**), *cat.* PdCl₂, dppf, Na₂CO₃, toluene, EtOH, H₂O. (d) DMF, 140 °C. (e) **i.** (*n*-Bu)₄NOH, THF. **ii.** HCl.

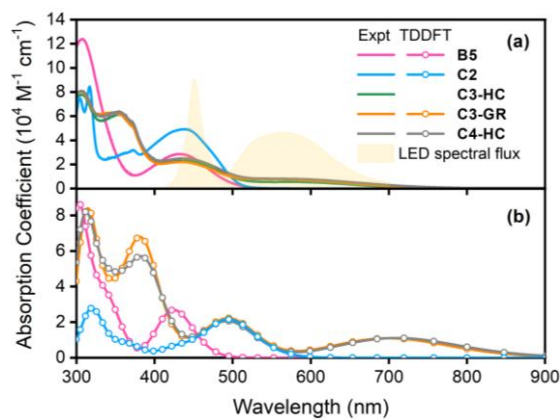

**Fig. 1** (a) Experimental and (b) simulated UV-Vis absorption spectra in THF for BfNI and BfNA derivatives.

**Table 1** Parameters for optoelectronic and electrochemical properties for BfNI derivatives.

| | $\lambda_{abs}$ ($\varepsilon \times 10^5$),[a] nm (M$^{-1}$ cm$^{-1}$) | $E_{ox}$ ($\Delta E_p$),[b] V vs Fc$^{0/+}$ | $E_{red}$ ($\Delta E_p$),[b] V vs Fc$^{0/+}$ | $E_{0-0}$,[c] eV | HOMO/LUMO,[d] V/V vs NHE |
|---|---|---|---|---|---|
| **B5** | 430 (0.77) | +0.26 | -1.98 | 2.48 | +0.89/-1.59 |
| **C2** | 373 (1.66), 434 (2.25) | +0.75[e] | -0.99 | 2.42 | +1.38/-1.04 |
| **C3-HC** | 357 (0.61), 436 (0.23), 565 (0.06) | +0.21 (0.10), +0.91[f] | -1.04 (0.09), -1.40 (0.14) | 1.65 | +0.84/-0.81 |
| **C3-GR** | 357 (0.79), 436 (0.28), 565 (0.10) | +0.24, +0.62 | -1.06 | 1.65 | +0.87/-0.78 |
| **C4-HC** | 354 (0.46), 434 (0.18), 563 (0.06) | +0.34 (0.16), +1.03[f] | -1.06 (0.16), -1.55 (0.19) | 1.60 | +0.97/-0.63 |

[a] Values were obtained in CH$_2$Cl$_2$. [b] Values were obtained in CH$_2$Cl$_2$. [c] Optical bandgap. [d] The HOMO energy level was determined from the oxidation potential measured by electrochemical methods, with ferrocene referenced to NHE at 0.63 V. The LUMO potential, representing the excited-state potential ($E_{0-0}{}^*$), was calculated using the HOMO energy and the optical bandgap according to the following equation: $E_{LUMO} = E_{HOMO} - E_{0-0}$. [e] Onset value. [f] Observed as irreversible peak.

highlight the effectiveness of unsymmetric extension in broadening the absorption spectrum reaching the NIR region.

Density functional calculations (at B3LYP and LC-BLYP level with $\omega = 0.14$ for ground state and time dependent cases, respectively) were performed to rationalize the electronic transitions. As shown in Figure 1(b) and Table S1-S3, the modelled methoxy derivatives **C3-GR** ($E_{cal} = 1.06$ eV, $f = 0.04$) and **C4-GR** ($E_{cal} = 0.91$ eV, $f = 0.03$) display lowest-energy transitions much lower than **B5** ($E_{cal} = 2.20$ eV, $f = 0.13$), indicating enhanced donor-acceptor interaction. Natural transition orbital (NTO) representation for the $S_0 \rightarrow S_1$ transition also shows clear charge-transfer character in both BfNI and BfNA derivatives compared to **B5**. Electron-hole separation (EHS) analysis reveals charge-transfer distances ($L$) of 9.31 Å (**C3-GR**) and 9.23 Å (**C4-GR**), significantly longer than that of **B5** (5.48 Å). These extended separations are consistent with enhanced ICT and NIR absorption.

Cyclic voltammetry further supports the redox-active nature of these BfNI and BfNA systems. As displayed in Figure 2, **B5** exhibits reversible redox couples at +0.26 V and −1.98 V (vs Fc/Fc$^+$). In contrast, **C2** exhibits a reversible reduction at −0.99 V and an irreversible oxidation onset at approximately +0.75 V, attributable to the absence of donor groups. Upon introduction of two triarylamine donors, **C3-HC** displays two reversible reduction peaks at −1.04 V and −1.40 V, along with one reversible oxidation at +0.21 V and one irreversible oxidation at +0.91 V. Similar electrochemical behavior is observed for both **C3-GR** and the BfNA analogue **C4-HC** ($E_{red}$ at −1.06 V and −1.55 V; $E_{ox}$ at +0.34 V and 1.03 V). These results clearly indicate the preservation of the inherent LUMO level of **C2** and the HOMO characteristics derived from the triarylamine donor (as in **B5**) in the target compound **C3-HC** (Fig. S1). The LUMO level is close to that for indole-fused (-0.88 V to -1.21 V vs Fc/Fc$^+$)[43] and thiophene-fused (-0.5 V vs Ag/AgCl)[44] analogues. In our case, we are able to convert BTD to BfNI/BfNA within 2 steps to achieve very low LUMO level (-1.59 to -0.81 V vs NHE). Although many reported fused-ring NI derivatives[12,13,40] and fused non-NI electron-deficient moieties, such as naphtho[1,2-c:5,6-c']bis[1,2,5]oxadiazoles (NOz2T),[45] exhibit similar redox potentials, they suffer from significant synthetic drawbacks. These include the need for multi-step procedures such as cross-coupling at NI core positions followed by cyclization, C−H activation, or oxidative coupling to obtain the desired products.[29] DFT-derived HOMO-LUMO energy levels further reflect the observed trends: from −4.68 eV/−2.21 eV for **B5** to −4.58 eV/−3.28 eV for **C3-GR** and −4.65 eV/−3.50 eV for **C4-GR**, highlighting the MO stabilizing effect of the electron-deficient BfNI/BfNA core. Orbital density plots show that the unoccupied frontier orbitals are delocalized across the entire BfNI/BfNA framework, indicative of strong π-conjugation and potential charge-transport properties. Electrostatic potential (ESP) mapping further reveals regions of high electron density localized at the fused imidazole and carboxyl groups, implying strong local polarization (Table S2). This planar and polarized architecture may support future exploration of BfNI/BfNA scaffolds in cation-binding or electrochemical storage applications.[19]

In summary, we present a concise and streamlined synthetic strategy for unsymmetric rylene imides, including benzimidazole-fused naphthalene imide (BfNI) and anhydride (BfNA) derivatives. This approach eliminates multistep protection-deprotection protocols, yielding π-extended chromophores with high solubility and strong, tunable absorption from the visible to near-infrared region. Fusion with electron-deficient benzimidazole lowers the LUMO, while triarylamine donors enhance redox stability and charge transfer, as supported by experiments and theoretical study. The resulting push-pull systems exhibit efficient charge management, tunable optoelectronic properties and energy-level alignment for device integration. Altogether, the structural simplicity, broadband absorption, and redox tunability of BfNI/BfNA derivatives highlight their potential for indoor photovoltaics and energy storage applications. Further investigations into their integration in energy-harvesting and storage devices are currently ongoing.

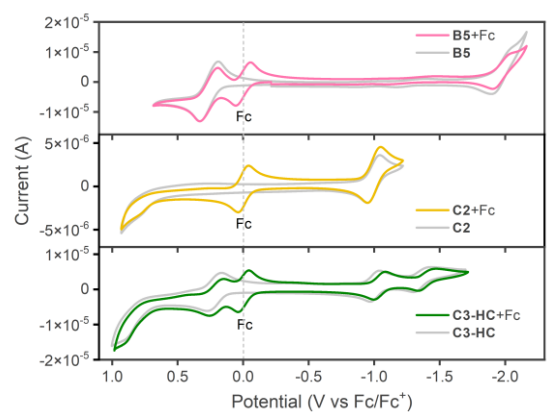

**Fig. 2** Cyclic voltammogram for compound B5, C2, and **C3-HC**.

This work was supported by the National Science and Technology Council (NSTC) in Taiwan (MOST 111-2113-M-126-002, 112-2113-M-126-002). Computational and storage resources were provided by the National Center for High-performance Computing (NCHC) of National Applied Research Laboratories (NARLabs).

## Conflicts of interest

There are no conflicts to declare.

## Data availability

The data supporting this article have been included as part of the SI. See DOI: https://doi.org/xxx

# Unsymmetric synthesis of panchromatic absorbing benzimidazole-fused naphthalene imides for multifunctional optoelectronic applications

Guan-Ru Lin,[a,‡] Huai-Chih Chang,[a,‡] Yi-Chen Wu,[a] Chen-Kai Hsieh,[a] Chih-Jou Chien,[a] Guan-Lin Lu,[a] Makeshmuralikrishna Kulasekaran,[b] Milanmathew Sssuraj,[c] Tzu-Ling Ho,[a] Jatin Rawat,[d] and Hsien-Hsin Chou*[a]

[a] Department of Applied Chemistry, Providence University, Taichung 433303, Taiwan

[b] Department of Chemistry, SRM Institute of Science and Technology, Kattankulathur, Tamil Nadu 603203, India

[c] Department of Chemistry, Karpagam Academy of Higher Education, Coimbatore, Tamil Nadu 641021, India

[d] Department of Chemical Engineering, Institute of Chemical Technology Mumbai – IndianOil Odisha Campus, Bhubaneswar, Odisha 751013, India

E-mail: hhchou@pu.edu.tw

**Experimental Methods**

**General information**

*General methods.* All manipulations were carried out under a nitrogen atmosphere using standard Schlenk techniques unless otherwise specified. Solvents were purified as follows: hexane, dichloromethane ($CH_2Cl_2$), and diethyl ether ($Et_2O$) were distilled from calcium hydride ($CaH_2$); tetrahydrofuran (THF) and toluene from sodium/benzophenone ketyl; and methanol from magnesium/iodine ($Mg/I_2$). All other solvents were reagent grade and used without further purification. Reagents were purchased from commercial suppliers and used as received unless stated otherwise. Tetra-*n*-butylammonium perchlorate (TBAP) was recrystallized from absolute ethanol and dried under vacuum prior to use. Thin-layer chromatography (TLC) was performed on silica gel GF254 plates (Agela Technologies), and column chromatography was carried out using 45–75 μm silica gel (Merck).

**Characterization**

*Spectrometric and spectroscopic characterization.* Mass spectra were acquired using a Finnigan TSQ Quantum Ultra EMR Triple Quadrupole LC/ESI/APCI mass spectrometer (Thermo Scientific). NMR spectra were recorded on a Bruker AVIII 400 HD spectrometer at room temperature and are reported in δ units, referenced to residual proton signals of the solvent ($CDCl_3$, δ = 7.26 ppm). UV-visible absorption spectra of $CH_2Cl_2$ solution containing each sample was recorded on a Jasco V-730 spectrophotometer, while photoluminescence (PL) spectra were obtained using a Jasco FP-8500 spectrofluorometer.

*Electrochemical characterization.* Cyclic voltammetry (CV) measurements were performed using a CH Instruments 6122E electrochemical workstation in a standard three-electrode configuration. A platinum disk electrode served as the working electrode, a platinum wire as the counter electrode, and a silver/silver nitrate ($Ag/AgNO_3$) electrode as the reference. Measurements were carried out in tetrahydrofuran solutions containing 0.1 M tetra-*n*-butylammonium perchlorate (TBAP) as the supporting electrolyte under a nitrogen atmosphere. The scan rate was typically 100 mV s$^{-1}$. Ferrocene/ferrocenium (Fc/Fc$^+$) was used as an internal standard, and redox potentials were calibrated against the normal hydrogen electrode (NHE) by referencing Fc/Fc$^+$ to 0.63 V vs NHE.

*Density functional calculations.* The ground-state geometries of the compounds were optimized using B3LYP with the 6-31G(d) basis set for all the molecules. Frequency calculations were carried out at the same level of theory as geometry optimization to confirm successful optimization of the geometries to stable points. Time-dependent density functional theory (TDDFT) calculations were performed on all molecules using the same basis sets and LC-BLYP as long-range-corrected (LRC) functional. The range-separation parameter ω were set to 0.15, 0.12, and 0.13 a.u. for **B5**, **C3-GR**, and **C4-GR**, respectively, according to the calculation of the situation where 0% Hartree-Fock exists in the short-range to probe the suitable short-range criteria for non-Coulomb part of exchange. All calculations were performed with the Gaussian 16[1] program package. The resultant vertical excitation profile is followed by simulation of absorption spectra using SWizard[2] software with a full width at half maximum of 3,000 cm[-1]. Natural transition orbitals (NTOs) and electron-hole separation (EHS) were analyzed with Multiwfn[3] program to compare the abilities of charge transfer.

### *Synthesis of building blocks for BfNI compounds*

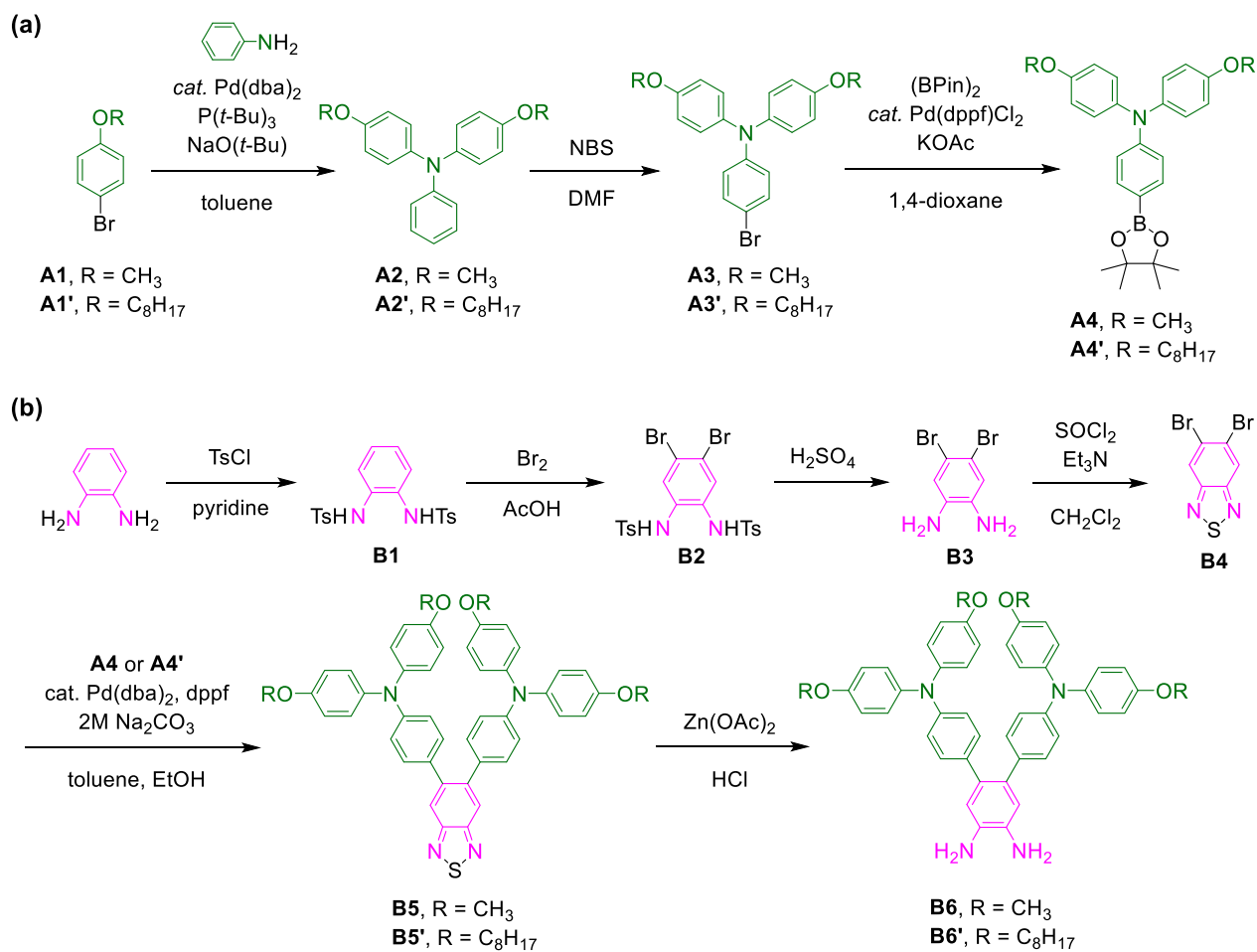

**Scheme S1** Synthetic routes for BfNI and BfNA derivatives.

***4-methoxy-N-(4-methoxyphenyl)-N-phenylaniline (A2).*** Under a nitrogen atmosphere, a three-necked flask containing a DMF solution of **A2** (3.00 g, 9.82 mmol in 21.5 mL) was treated dropwise with a DMF solution of N-bromosuccinimide (NBS) (2.10 g, 11.78 mmol in 21.5 mL) at room temperature., Pd(dba)$_2$ (0.944 g, 1.64 mmol) and sodium tert-butoxide NaO($t$-Bu), 9.85 g, 102.5 mol) were weighted into the double-necked round-bottom flask, followed by the sequential addition of aniline (3.0 mL, 32.86 mmol), compound **A1** (9.0 mL, 72.29 mmol), tri($t$-butyl)phosphine (1.31 mL of a 0.48 M solution in toluene, 2.63 mmol), and toluene (200 mL). The reaction mixture was heated to 100 °C and stirred for 18 hours. After cooling to room temperature, the mixture was concentrated under reduced pressure and extracted with H$_2$O and CH$_2$Cl$_2$. The organic phase was dried over anhydrous MgSO$_4$, filtered, and concentrated to give a black solid. Purification by silica gel column chromatography (hexane/CH$_2$Cl$_2$, 8:2 v/v) afforded compound **A2** as a white powder (8.89 g, 89% yield). $^1$H NMR (CDCl$_3$): δ 7.19 (t, J = 8.8 Hz, 2H), 7.07 (d, J = 9.2 Hz, 4H), 6.97 (d, J = 8.8 Hz, 2H), 6.90–6.83 (m, 5H), 3.82 (s, 6H).

Compound **A2'** was synthesized following the same procedure as for **A2**, starting from **A1'** (1.02 g). The product was obtained as a white oil (3.06 g, 56% yield).$^1$H NMR (CDCl$_3$): δ 7.17-7.04 (m, 6H), 6.96 (d, J = 8.4 Hz, 2H), 6.83 (d, J = 8.0 Hz, 6H), 3.94 (s, 4H), 1.82-0.89 (m, 63H).

***4-bromo-N,N-bis(4-methoxyphenyl)aniline (A3).*** Under a nitrogen atmosphere, a three-necked flask containing a DMF solution of **A2** (3.00 g, 9.82 mmol in 21.5 mL) was treated dropwise with a DMF solution of *N*-bromosuccinimide (2.10 g, 11.78 mmol in 21.5 mL) at room temperature. The reaction was stirred for 18 hours. After completion, the mixture was extracted with H$_2$O and CH$_2$Cl$_2$, and the organic layer was dried over anhydrous MgSO$_4$, filtered, and concentrated under reduced pressure to yield a purple solid. Purification by flash chromatography on silica gel (hexane/CH$_2$Cl$_2$, 8:2 v/v) afforded **A3** as a white powder (2.02 g, 53% yield). $^1$H NMR (CDCl$_3$): δ 7.26 (d, J = 8.8 Hz, 2H), 7.05 (d, J = 9.2 Hz, 4H), 6.86 (d, J = 5.6 Hz, 2H), 6.83 (d, J = 6.8 Hz, 2H), 3.81 (s, 6H).

Compound **A3'** was synthesized following the same procedure as for **A3**, starting from **A2'** (3.06 g). The product was obtained as a white oil (0.83 g, 24% yield). $^1$H NMR (CDCl$_3$): δ 7.24 (d, J = 9.2 Hz, 2H), 7.03 (d, J = 8.8 Hz, 4H), 6.85-6.79 (m, 6H), 3.94 (t, J = 13.2 Hz, 4H), 1.86-0.89 (m, 63H).

***4-methoxy-N-(4-methoxyphenyl)-N-(4-(4,4,5,5-tetramethyl-1,3,2-dioxaborolan-2-yl)phenyl)aniline (A4).*** Compound **A3** (1.90 g, 4.95 mmol), bis(pinacolato)diboron (1.64 g, 6.47 mmol), PdCl$_2$ (0.09 g, 0.50 mmol), 1,1'-bis(diphenylphosphino)ferrocene (DPPF, 0.29 g, 0.50 mmol), and potassium acetate (1.46 g, 14.89 mmol) were weighed into a double-necked flask. 1,4-Dioxane (50 mL) was added via syringe after the flask had been flushed with nitrogen. The reaction mixture was heated to 95 °C and stirred for 24 hours. After cooling, the mixture was extracted with ethyl acetate and washed with water. The organic layer was dried over anhydrous MgSO$_4$, filtered, and concentrated under reduced pressure to afford a dark brown liquid (2.01 g). The crude product, assigned as **A4**, was used directly in the next step without further purification. [1]H NMR (CDCl$_3$): δ 7.61 (d, J = 8.0 Hz, 2H), 7.08 (d, J = 9.2 Hz, 4H), 6.88 (d, J = 8.4 Hz, 2H), 6.85 (d, J = 9.2 Hz, 2H), 3.82 (s, 6H), 1.34 (s,12H).

Compound **A4'** was synthesized following the same procedure as for **A4**, starting from **A3'** (0.83 g). The product was obtained as a white oil (0.44 g, 41% yield). [1]H NMR (CDCl$_3$): δ 7.61 (d, J = 8.0 Hz, 2H), 7.07 (d, J = 8.0 Hz, 4H), 6.88 (d, J = 8.0 Hz, 2H), 6.84 (d, J = 8.0 Hz, 4H), 3.73 (d, J = 1.2 Hz, 4H), 1.18-0.86 (m, 169H).

***N,N'-(1,2-Phenylene)bis(4-methylbenzenesulfonamide) (B1).*** In a double-necked flask, *o*-phenylenediamine (10.00 g, 92.51 mmol) and *p*-toluenesulfonyl chloride (35.30 g, 185.01 mmol) were added under a nitrogen atmosphere. The flask was cooled in an ice bath, followed by the addition of pyridine (100 mL). The reaction mixture was stirred at room temperature for 24 hours. Upon completion, 15% aqueous HCl was added, and the resulting precipitate was collected by vacuum filtration. Recrystallization from ethanol afforded **B1** as a white powder (34.68 g, 90% yield). [1]H NMR (CDCl$_3$): δ 7.60 (d, J = 8.0 Hz, 4H), 7.25 (d, J = 8.0 Hz, 4H), 7.07 (dt, J = 7.4, 3.6 Hz, 2H), 7.06–6.98 (m, 2H), 6.80 (s, 2H), 2.42 (s, 6H)..

***N,N'-(4,5-dibromo-1,2-phenylene)bis(4-methylbenzenesulfonamide) (B2).*** In a three-necked flask, **B1** (5.02 g, 12.00 mmol) and sodium acetate (2.17 g, 26.41 mmol) were combined, followed by the addition of acetic acid (50 mL) under cooling in an ice bath. Bromine (1.42 mL, 8.09 mmol) was then added dropwise while maintaining the temperature. After the addition was complete, the reaction mixture was heated to 110 °C and stirred for 3 hours. The resulting solution was poured into cold water (50 mL) and stirred vigorously for 1 hour. Ethanol (25 mL) was then added, and the mixture was stirred for an additional 30 minutes. The resulting precipitate was collected by vacuum filtration and washed thoroughly with water to afford **B2** as a white powder (5.51 g, 79% yield).[1]H NMR (CDCl$_3$): δ 7.62 (d, J = 8.4 Hz, 4H), 7.31 (d, J = 8.0 Hz, 4H), 7.23 (s, 2H), 6.80 (s, 2H), 2.45 (s, 6H).

***4,5-dibromobenzene-1,2-diamine (B3).*** Under a nitrogen atmosphere, **B2** (1.0054 g, 1.74 mmol) was weighed into a double-necked flask, followed by the addition of concentrated sulfuric acid while maintaining the mixture in an ice bath. The reaction was then heated to 110 °C and stirred for 15 minutes before cooling to room temperature. Cold water (10 mL) was added, followed by dropwise addition of 50% aqueous NaOH until the pH reached neutral. The resulting precipitate was collected by filtration to afford **B3** as a pink powder (0.32 g, 71% yield). $^1$H NMR (CDCl$_3$): δ 6.95 (s, 2H). 3.43 (br, 4H).

***Synthesis of BfNI- and BfNA-series compounds.***

***5,6-dibromobenzo[c][1,2,5]thiadiazole (B4).*** Under a nitrogen atmosphere, **B3** (0.5014 g, 1.91 mmol) was placed in a three-necked flask. CH$_2$Cl$_2$ (15 mL) and Et$_3$N (7.5 mL) were added sequentially, followed by the slow addition of thionyl chloride (0.7 mL, 3.48 mmol) under ice-cooling. After removal of the ice bath, the reaction mixture was heated to 60 °C and refluxed for 2 hours. Upon completion, the mixture was concentrated and extracted with water and CH$_2$Cl$_2$. The organic layer was dried over anhydrous MgSO$_4$ and filtered through Celite. Volatiles were removed under reduced pressure to yield a black residue, which was purified by column chromatography on silica gel (hexane as eluent) to afford **B4** as a white powder (0.49 g, 88% yield). $^1$H NMR (CDCl$_3$): δ 8.42 (s, 2H). $^{13}$C NMR (CDCl$_3$): δ 153.8, 127.3, 125.0.

***4,4'-(benzo[c][1,2,5]thiadiazole-5,6-diyl)bis(N,N-bis(4-methoxyphenyl)aniline) (B5).*** Under a nitrogen atmosphere, compound **B4** (0.18 g, 0.61 mmol), **A4** (1.05 g, 2.44 mmol), PdCl$_2$ (0.02 g, 0.13 mmol), and DPPF (0.07 g, 0.12 mmol) were weighed into a double-necked flask. Toluene (6 mL), ethanol (0.7 mL), and 2 M aqueous Na$_2$CO$_3$ (1.45 mL) were then added sequentially. The reaction mixture was heated to 130 °C and stirred for 24 hours. After cooling to room temperature, the mixture was extracted with H$_2$O and CH$_2$Cl$_2$. The organic layer was dried over anhydrous MgSO$_4$, filtered, and concentrated under reduced pressure to yield a dark brown residue. Purification by column chromatography on silica gel using hexane/CH$_2$Cl$_2$ (7:3, v/v) as the eluent afforded **B5** as an orange powder (0.31 g, 68% yield). $^1$H NMR (CDCl$_3$): δ 7.99 (s, 2H), 7.10-7.07 (m, 12H), 6.88-6.86 (m, 12H), 3.82 (s, 12H). $^{13}$C NMR (CDCl$_3$): δ 156.0, 154.4, 144.0, 140.7, 130.4, 126.7, 120.9, 119.5, 114.7, 55.5. HR-MS calcd. for C$_{46}$H$_{39}$O$_4$N$_4$S: 743.2687; Found: 743.2742 (M$^+$).

Compound **B5'** was synthesized following the same procedure as for **B5**, starting from **B4'** (0.06 g). The product was obtained as a white oil (0.07 g, 89% yield). $^1$H NMR (CDCl$_3$): δ 7.95 (s, 2H), 7.26 (d, J = 2.8 Hz, 4H) 6.98 (d, J = 8.8 Hz, 2H), 6.84-6.81 (m, 12H), 3.92 (t, J = 6.8 Hz, 6H), 1.81-1.73 (m, 8H), 1.47-1.29 (m,

42H), 0.89 (t, J = 5.1 Hz, 12H). $^{13}$C NMR (CDCl$_3$): δ 155.6, 154.4, 148.1, 144.0, 140.5, 132.0, 130.4, 126.7, 120.9, 119.3, 115.3, 68.3, 31.9, 31.6, 29.4, 29.3, 26.1, 22.7, 14.1. HR-MS calcd. for C$_{74}$H$_{94}$O$_4$N$_4$S: 1134.6990; Found: 1134.7034 (M$^+$).

***N4,N4,N4'',N4''-tetrakis(4-methoxyphenyl)-4',5'-dihydro-[1,1':2',1''-terphenyl]-4,4',4'',5'-tetraamine (B6).*** Under a nitrogen atmosphere, **B5** (0.28 g, 0.38 mmol) and zinc powder (0.39 g, 5.97 mmol) were suspended in acetic acid (4 mL). The mixture was heated to 75 °C and stirred for 12 hours. Upon completion, the reaction mixture was filtered through Celite under nitrogen, and the volatiles were removed under reduced pressure to afford **B6** as a gray solid. Due to its air sensitivity, the crude product was used directly in the subsequent step without further purification.

Compound **B6'** was synthesized following the same procedure as for **B6**, starting from **B5'** (0.43 g). The product was obtained as a white oil.

***7-hexadecyl-1H-isochromeno[6,5,4-def]isoquinoline-1,3,6,8(7H)-tetraone (C1).*** Naphthalene tetracarboxylic dianhydride (NTCDA, 1.00 g, 3.73 mmol) was placed in a three-necked flask, and hexadecylamine (0.29 g, 1.25 mmol) was loaded into a dropping funnel. Under a nitrogen atmosphere, *N*,*N*-dimethylformamide (70 mL) was added in two equal portions to the flask and funnel, and the amine solution was added dropwise. The resulting mixture was then heated to 140 °C and stirred for 24 hours. After completion, water was added to precipitate the product, which was collected by filtration. The solid was extracted with H$_2$O and CH$_2$Cl$_2$, and the organic layer was dried over anhydrous MgSO$_4$. After filtration and solvent removal under reduced pressure, the desired product was obtained as a beige solid (0.42 g, 67% yield).

***9,10-dibromo-2-hexadecylbenzo[lmn]benzo[4,5]imidazo[2,1-b][3,8]phenanthroline-1,3,6(2H)-trione (C2).*** Compound **C1** (1.0003 g, 2.03 mmol) and **B3** (1.1039 g, 4.15 mmol) were placed in a double-necked flask. After sealing and purging with nitrogen, DMF (14 mL) was added. The resulting mixture was stirred and heated to 140 °C for 24 hours. Upon completion, the reaction mixture was poured into water and filtered to remove DMF. The crude residue was then treated with a MeOH/THF mixture to induce precipitation. The solid was collected by filtration to afford an orange-red powder (0.76 g, 52% yield). $^1$H NMR (CDCl$_3$): δ 8.98 (d, J = 7.6 Hz, 1H), 8.95 (d, J 7.6 Hz, 1H), 8.87 (s, 1H), 8.85 (d, J = 7.6 Hz, 1H), 8.82 (d, J = 7.6 Hz, 1H), 8.20 (s, 1H), 4.21 (t, J = 8.0 Hz, 2H), 1.76-1.25 (m, H), 0.87 (t, J = 6.8 Hz, 3H). HR-MS calcd. for C$_{36}$H$_{39}$O$_3$N$_3$Br$_2$: 719.1353; Found: 719.1343 (M$^+$).

***9,10-bis(4-(bis(4-(octyloxy)phenyl)amino)phenyl)-2-hexadecylbenzo[lmn]benzo[4,5]imidazo[2,1-b][3,8]phenanthroline-1,3,6(2H)-trione (C3-HC).*** Compound **C2** (204.0 mg, 282.7 μmol), **A4'** (673.0 mg, 1072.1 μmol), PdCl$_2$ (14.9 mg, 84.0 μmol), and DPPF (40.5 mg, 73.1 μmol) were placed in a two-necked flask and evacuated. Toluene (3.0 mL) was added under a nitrogen atmosphere, and the mixture was stirred at 80 °C. Subsequently, ethanol (0.35 mL) and 2 M aqueous Na$_2$CO$_3$ (0.693 mL, 138.6 μmol) were added, and the reaction temperature was raised to 130 °C and maintained for 24 hours. After cooling, the mixture was extracted with H$_2$O and CH$_2$Cl$_2$. The organic layer was dried over anhydrous MgSO$_4$, filtered, and concentrated under reduced pressure to yield a dark green residue. Purification by column chromatography (hexane/CH$_2$Cl$_2$, 2:8 v/v) afforded a dark green solid (31.7 mg, 7% yield). [1]H NMR (CDCl$_3$): δ 8.79 (dd J = 8.0, 16 Hz, 2H), 8.70 (dd, J = 8.0, 16 Hz, 1H), 8.42 (s, 1H), 7.81 (s, 1H), 7.19-7.00 (m, 12H), 6.94-6.8 (m, 12H), 4.18 (t, J = 8.0 Hz, 2H), 4.00 (s, 8H), 2.17-1.68 (m, 25H), 1.68-1.11 (m, 113H), 1.11-0.65 (m, 30H) 0.12 (s, 1H). [13]C NMR (CDCl$_3$)s: δ 162.77, 162.63, 158.87, 155.70, 155.46, 147.98, 147.70, 147.60, 147.52, 143.20, 142.84, 141.45, 140.73, 140.47, 139.95, 133.24, 133.12, 132.75, 132.67, 131.58, 130.95, 130.78, 130.66, 130.21, 130.12, 127.94, 127.57, 127.32, 126.88, 126.59, 125.67, 125.43, 124.51, 124.38, 122.19, 121.73, 121.61, 120.29, 119.77, 117.35, 116.93, 116.83, 115.28, 77.38, 77.06, 76.74, 68.30, 40.98, 31.96, 31.87, 29.74, 29.60, 29.44, 29.31, 28.14, 27.16, 26.15, 22.71, 14.15, 1.06. HR-MS calcd. for C$_{104}$H$_{133}$O$_7$N$_5$: 1564.0200; Found: 1564.0162 (M$^+$).

An alternative route (Method B) was employed as follows: **C1** (14.5 mg, 29.5 μmol) and **B6'** (88.8 mg, 80.2 μmol) were weighted into a two-necked flask under nitrogen atmosphere. DMF (0.3 mL) was then syringed into the flask. The reaction mixture was stirred at 140 °C in the dark for 24 hours. Upon completion, the mixture was extracted with H$_2$O and CH$_2$Cl$_2$. The organic layer was dried over anhydrous MgSO$_4$, filtered, and concentrated under reduced pressure to give a black liquid. Purification by column chromatography (hexane/CH$_2$Cl$_2$, 2:8 v/v) afforded a dark green solid (28.1 mg, 61% yield).

Compound **C3-GR** was synthesized following the same procedure (Method B) as for **C3-HC**, starting from **C1** (60.0 mg, 122.0 μmol) and **B6** (270.0 mg, 377.7 μmol), affording the product as a white oil (45.4 mg, 32% yield). [1]H NMR (CDCl$_3$): δ 9.02 (d, J = 8.0 Hz, 1H), 8.91 (d, J = 8.0 Hz, 1H), 8.81 (dd, 2H), 8.53 (s, 1H), 7.92 (s, 1H), 7.08-7.04 (m, 12H), 6.88-6.86 (m, 12H), 4.22 (t, J = 8.4 Hz, 2H), 3.83 (s, 12H), 1.85-1.71 (m, 4H), 1.45-1.28 (m, 24H), 0.90 (t, J = 6.8, 3H). [13]C NMR (CDCl$_3$): δ 162.7, 162.6, 158.9, 155.8, 147.9, 147.5, 147.4, 142.3, 140.9, 140.7, 140.0, 133.3, 133.2, 131.6, 131.2, 130.9, 130.8, 130.6, 127.7, 127.3, 127.2,

126.8, 126.5, 125.7, 125.1, 124.6, 121.4, 119.9, 116.9, 114.7, 55.5, 53.4, 41.0, 31.9, 29.7, 29.6, 29.5, 29.3, 22.7, 14.1. HR-MS calcd. For $C_{76}H_{76}O_7N_5$: 1170.5739; Found: 1170.5717 (M$^+$).

***9,10-bis(4-(bis(4-(octyloxy)phenyl)amino)phenyl)-1H-benzo[4,5]imidazo[2,1-a]isochromeno[6,5,4-def]isoquinoline-1,3,6-trione (C4-HC).*** Using Method A: compound **C3-HC** (10.0 mg, 6.4 μmol) was placed in a double-necked flask, flushed with nitrogen, and dissolved in THF (1 mL). Tetra-*n*-butylammonium hydroxide (15 μL, 1 M in MeOH, 15 μmol) was added under a nitrogen atmosphere. The reaction mixture was stirred at room temperature for 24 hours in the dark. After completion, the mixture was extracted with $H_2O$ and $CH_2Cl_2$. The organic phase was dried over anhydrous $MgSO_4$, filtered, and concentrated under reduced pressure to afford a green solid. The crude product was further precipitated from MeOH/DCM, filtered, and collected as **C4-HC**, a green solid (3.6 mg, yield 42%). $^1$H NMR (CDCl$_3$): δ 9.12 (d, J = 8.0 Hz, 1H), 8.96 (d, J = 8.0 Hz, 1H), 8.85 (dd, 2H), 8.57 (s, 1H), 7.97 (s, 1H), 7.22-7.00 (m, 12H), 6.95-6.70 (m, 12H), 4.24 (t, J = 8.0 Hz, 2H), 4.00 (s, 8H), 1.85-1.61 (m, 45H), 1.53-1.22 (m, 72H), 0.90 (t, J = 8, 12H). $^{13}$C NMR (CDCl$_3$): δ 162.98, 162.86, 159.16, 155.51, 148.35, 148.11, 147.94, 147.73, 147.54, 143.22, 141.85, 140.72, 139.99, 133.30, 133.13, 132.85, 131.69, 131.22, 130.88, 130.75, 130.65, 130 30, 130.19, 128.08, 127.98, 127.70, 127.49, 126.66, 125.92, 125.75, 124.51, 122.87, 121.77, 120.69, 120.44, 119.82, 116.99, 115.32, 77.40, 77.08, 76.77, 68 34, 41.05, 31.93, 29.73, 29.43, 28.19, 27.18, 26.17, 22.74, 14.18, 1.09. HR-MS calcd. for $C_{88}H_{99}O_8N_4$: 1339.74574; Found: 1339.74623 (M$^+$).

Alternatively, compound **C4-HC** was synthesized following a similar procedure to that of **C3-HC** (Method B), using NTCDA and B6' as starting materials. However, the product was obtained only in trace amounts and with considerable impurities.

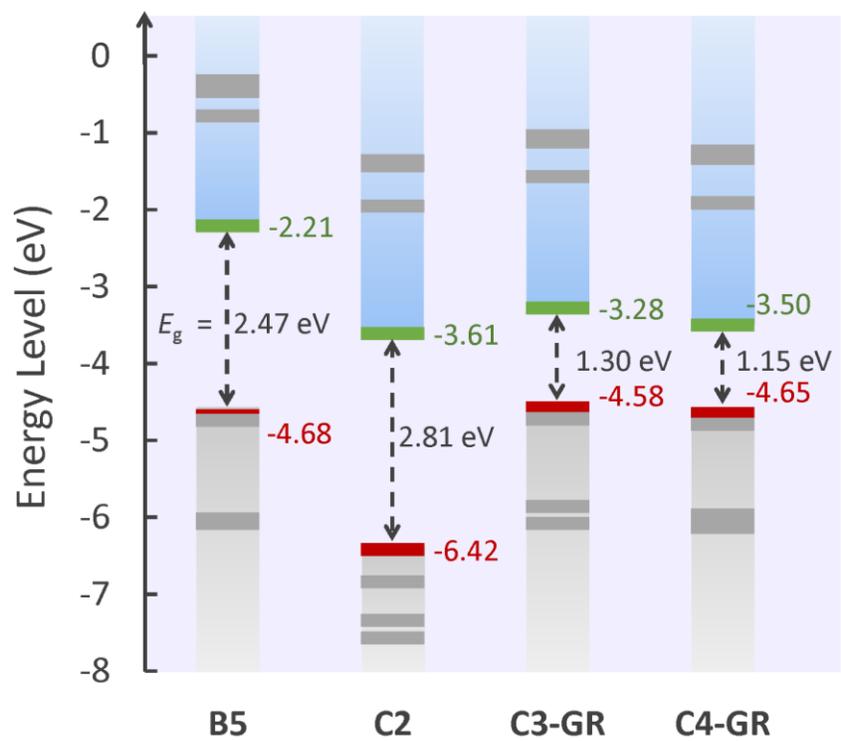

**Fig. S7** Energy level diagram for compound **B5**, **C2**, **C3-GR**, and **C4-GR**

**Table S1** Natural transition orbitals (NTO) and electron−hole separations (EHS) for the first excited states in GR-series compounds.[a\]

| Dye | $E_{eb}$ (eV) | NTO Hole | $E_{cal}$ (population) | Electron | EHS $L$ [Å] ($t$ [Å]) |
|---|---|---|---|---|---|
| **B5** | 2.58 | 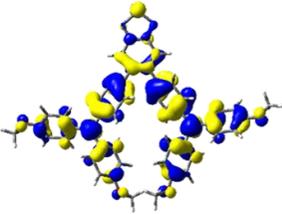 | 2.20 (99.7%) | 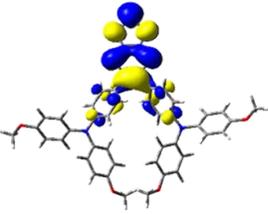 | 5.48 (2.80) |
| **C3-GR** | 1.69 | 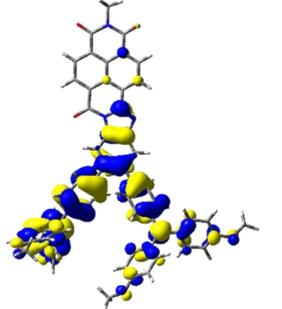 | 1.06 (99.7%) | 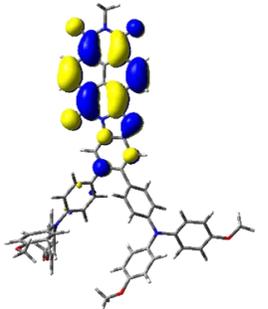 | 9.31 (6.24) |
| **C4-GR** | 1.72 | 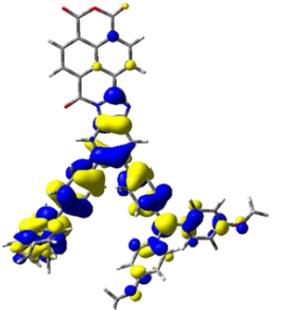 | 0.91 (99.7%) | 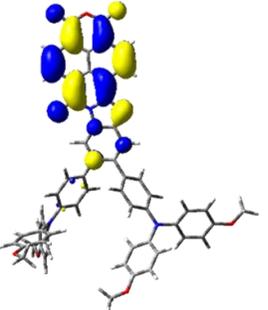 | 9.23 (6.15) |

[a] The results were computed at the TD-LC-BLYP/6-31G* level of theory and the orbitals were plotted with isovalue = 0.05. [b] $E_{cal}$ denotes to the lowest-energy vertical excitations. c The NTOs shown account for >99% of the respective total excitation process. [d] $E_{eb}$ denotes to exciton binding energy. [e] Those orbitals colored in blue represent depletion of electron density, while that in orange represent increment of electron density. $L$ represents the charge transfer distance. The $t$ index represents the degree of electron−hole separation.

**Table S2** Molecular geometries and corresponding electrostatic potential (ESP) maps for the BfNI dyes.[a]

| | Molecular geometry | Neutral ESP | Monocationic ESP |
|---|---|---|---|
| | | − + | − + |
| **B5** | 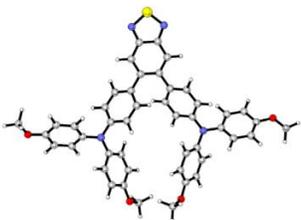 | 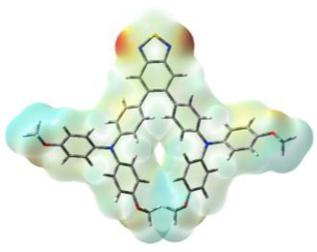 | 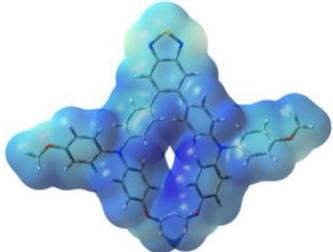 |
| **C3-GR** | 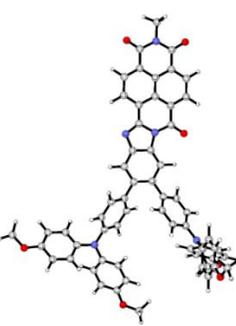 | 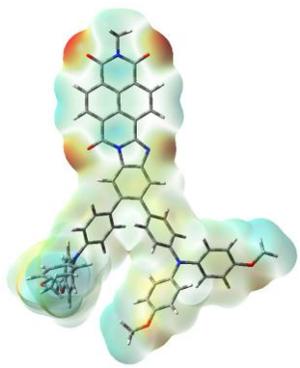 | 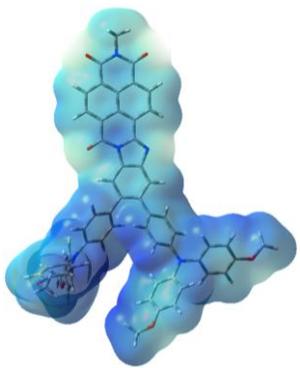 |
| **C4-GR** | 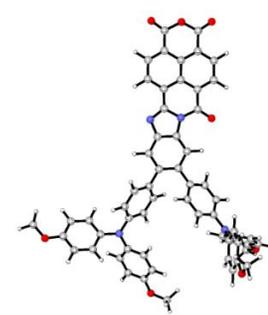 | 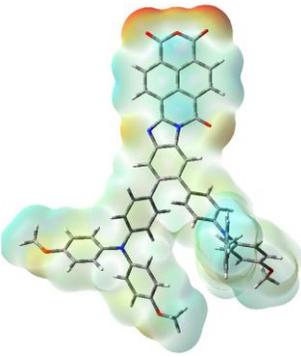 | 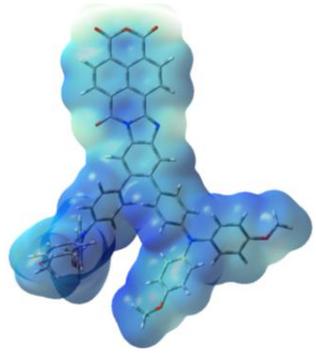 |

| B5 | | | | | | | | |
|---|---|---|---|---|---|---|---|---|
| Transition | Ecal (eV) | f | Composition | | Transition | Ecal (eV) | f | Composition | |
| S1 | 2.20 | 0.13 | H→L | 98% | S9 | 3.64 | 0.00 | H4→L | 35% |
| S2 | 2.22 | 0.04 | H1→L | 98% | | | | H3→L | 37% |
| S3 | 3.42 | 0.22 | H→L1 | 96% | | | | H2→L | 14% |
| S4 | 3.53 | 0.01 | H4→L | 55% | S10 | 3.66 | 0.02 | H3→L | 6% |
| | | | H3→L | 16% | | | | H1→L2 | 56% |
| | | | H2→L | 27% | | | | H→L3 | 27% |
| S5 | 3.57 | 0.26 | H1→L1 | 96% | S11 | 3.95 | 0.00 | H1→L2 | 29% |
| S6 | 3.61 | 0.06 | H5→L | 25% | | | | H→L3 | 64% |
| | | | H4→L | 6% | S12 | 3.98 | 0.04 | H1→L3 | 72% |
| | | | H3→L | 11% | | | | H→L2 | 14% |
| | | | H2→L | 38% | S13 | 4.01 | 0.11 | H6→L | 38% |
| | | | H→L2 | 11% | | | | H→L4 | 54% |
| S7 | 3.62 | 0.11 | H5→L | 43% | S14 | 4.04 | 0.01 | H6→L | 47% |
| | | | H1→L3 | 9% | | | | H1→L4 | 6% |
| | | | H→L2 | 41% | | | | H→L4 | 35% |
| S8 | 3.63 | 0.10 | H5→L | 19% | S15 | 4.05 | 0.02 | H1→L4 | 76% |
| | | | H3→L | 27% | | | | H→L5 | 6% |
| | | | H2→L | 19% | | | | | |
| | | | H→L2 | 27% | | | | | |

| C3-GR | | | | | | | | |
|---|---|---|---|---|---|---|---|---|
| Transition | Ecal (eV) | f | Composition | | Transition | Ecal (eV) | f | Composition | |
| S1 | 1.06 | 0.04 | H→L | 99% | S26 | 3.63 | 0.06 | H1→L5 | 29% |
| S2 | 1.21 | 0.07 | H1→L | 100% | | | | H→L6 | 25% |
| S3 | 2.25 | 0.20 | H2→L | 97% | | | | H→L7 | 33% |
| S4 | 2.56 | 0.14 | H5→L | 91% | S27 | 3.64 | 0.01 | H16→L | 89% |
| S5 | 2.60 | 0.00 | H3→L | 99% | S28 | 3.77 | 0.63 | H2→L1 | 21% |
| S6 | 2.63 | 0.00 | H5→L | 3% | | | | H1→L4 | 52% |
| | | | H4→L | 96% | | | | H→L6 | 17% |
| S7 | 2.73 | 0.21 | H1→L | 97% | S29 | 3.78 | 0.00 | H21→L | 87% |
| S8 | 2.86 | 0.02 | H1→L1 | 98% | S30 | 3.82 | 0.05 | H2→L1 | 8% |
| S9 | 2.95 | 0.07 | H6→L | 90% | | | | H1→L4 | 32% |
| S10 | 3.07 | 0.06 | H7→L | 96% | | | | H1→L5 | 6% |
| S11 | 3.11 | 0.00 | H11→L | 6% | | | | H→L6 | 24% |
| | | | H9→L | 8% | | | | H→L7 | 20% |
| | | | H8→L | 72% | S31 | 3.87 | 0.06 | H17→L | 42% |
| S12 | 3.15 | 0.10 | H→L2 | 84% | | | | H2→L1 | 21% |
| | | | | | | | | H→L6 | 14% |

| Transition | Ecal (eV) | f | Composition | | Transition | Ecal (eV) | f | Composition | |
|---|---|---|---|---|---|---|---|---|---|
| | | | H→ L3 | 6% | S32 | 3.91 | 0.08 | H17→ L | 9% |
| S13 | 3.22 | 0.02 | H11→ L | 45% | | | | H1→ L5 | 22% |
| | | | H10→ L | 16% | | | | H1→ L6 | 28% |
| | | | H9→ L | 33% | | | | H→ L5 | 19% |
| S14 | 3.24 | 0.00 | H15→ L | 96% | | | | H→ L6 | 10% |
| S15 | 3.28 | 0.00 | H11→ L | 14% | S33 | 3.94 | 0.03 | H17→ L | 19% |
| | | | H10→ L | 10% | | | | H2→ L1 | 21% |
| | | | H9→ L | 56% | | | | H1→ L5 | 7% |
| | | | H8→ L | 17% | | | | H1→ L6 | 36% |
| S16 | 3.28 | 0.01 | H→ L2 | 7% | S34 | 3.94 | 0.04 | H17→ L | 7% |
| | | | H→ L3 | 89% | | | | H2→ L1 | 17% |
| S17 | 3.31 | 0.09 | H11→ L | 13% | | | | H1→ L5 | 20% |
| | | | H10→ L | 33% | | | | H1→ L6 | 16% |
| | | | H1→ L2 | 47% | | | | H→ L7 | 22% |
| S18 | 3.31 | 0.04 | H11→ L | 15% | S35 | 3.99 | 0.00 | H22→ L | 84% |
| | | | H10→ L | 38% | | | | H20→ L | 6% |
| | | | H1→ L2 | 42% | S36 | 4.00 | 0.01 | H1→ L7 | 77% |
| S19 | 3.43 | 0.01 | H→ L2 | 6% | | | | H→ L7 | 7% |
| | | | H1→ L3 | 91% | S37 | 4.11 | 0.20 | H1→ L8 | 35% |
| S20 | 3.46 | 0.00 | H12→ L | 98% | | | | H→ L8 | 50% |
| S21 | 3.50 | 0.00 | H13→ L | 98% | | | | H→ L9 | 7% |
| S22 | 3.53 | 0.16 | H14→ L | 81% | S38 | 4.12 | 0.30 | H1→ L9 | 15% |
| | | | H→ L4 | 9% | | | | H→ L9 | 64% |
| S23 | 3.55 | 0.00 | H18→ L | 89% | S39 | 4.14 | 0.03 | H1→ L10 | 42% |
| S24 | 3.58 | 0.11 | H14→ L | 6% | | | | H→ L10 | 51% |
| | | | H1→ L7 | 6% | S40 | 4.14 | 0.01 | H1→ L13 | 11% |
| | | | H→ L4 | 57% | | | | H→ L11 | 15% |
| | | | H→ L5 | 18% | | | | H→ L12 | 21% |
| S25 | 3.62 | 0.08 | H1→ L5 | 6% | | | | H→ L13 | 39% |
| | | | H1→ L6 | 11% | | | | | |
| | | | H→ L4 | 21% | | | | | |
| | | | H→ L5 | 47% | | | | | |

## C4-GR

| Transition | Ecal (eV) | f | Composition | | Transition | Ecal (eV) | f | Composition | |
|---|---|---|---|---|---|---|---|---|---|
| S1 | 0.91 | 0.03 | H→ L | 99% | S25 | 3.61 | 0.09 | H2→ L1 | 8% |
| S2 | 1.06 | 0.08 | H1→ L | 99% | | | | H1→ L7 | 10% |
| S3 | 2.15 | 0.15 | H2→ L | 97% | | | | H→ L5 | 19% |
| S4 | 2.42 | 0.00 | H3→ L | 99% | | | | H→ L6 | 36% |
| S5 | 2.44 | 0.02 | H5→ L | 65% | | | | H→ L7 | 16% |
| | | | H→ L1 | 27% | S26 | 3.63 | 0.03 | H1→ L6 | 39% |
| S6 | 2.47 | 0.00 | H4→ L | 94% | | | | H→ L5 | 7% |
| S7 | 2.49 | 0.28 | H5→ L | 28% | | | | H→ L6 | 14% |
| | | | H→L1 | 68% | | | | H→L7 | 31% |
| S8 | 2.61 | 0.02 | H1→L1 | 99% | S27 | 3.66 | 0.59 | H2→L1 | 49% |
| S9 | 2.82 | 0.05 | H6→L | 92% | | | | H1→L5 | 7% |
| S10 | 2.93 | 0.07 | H7→L | 96% | | | | H→L5 | 12% |
| S11 | 2.97 | 0.00 | H11→L | 14% | | | | H→L6 | 9% |
| | | | H9→L | 52% | | | | H→L7 | 9% |
| | | | H8→L | 22% | S28 | 3.74 | 0.07 | H16→L | 11% |
| S12 | 3.05 | 0.05 | H→L2 | 76% | | | | H1→L4 | 33% |
| | | | H→L3 | 13% | | | | H→L5 | 29% |
| S13 | 3.07 | 0.02 | H11→L | 38% | | | | H→L7 | 8% |
| | | | H9→L | 35% | S29 | 3.76 | 0.05 | H16→L | 52% |
| | | | H8→L | 12% | | | | H1→L4 | 26% |
| | | | H→L2 | 6% | S30 | 3.78 | 0.00 | H16→L | 9% |
| S14 | 3.11 | 0.00 | H11→L | 25% | | | | H2→L1 | 25% |
| | | | H8→L | 64% | | | | H1→L4 | 27% |
| S15 | 3.13 | 0.03 | H→L2 | 14% | | | | H→L5 | 23% |
| | | | H→L3 | 82% | S31 | 3.85 | 0.05 | H1→L5 | 85% |
| S16 | 3.15 | 0.01 | H11→L | 12% | S32 | 3.91 | 0.01 | H1→L6 | 45% |
| | | | H10→L | 87% | | | | H1→L7 | 5% |
| S17 | 3.20 | 0.06 | H1→L2 | 89% | | | | H→L6 | 29% |
| | | | H1→L3 | 8% | | | | H→L7 | 17% |
| S18 | 3.28 | 0.04 | H12→L | 8% | S33 | 3.92 | 0.00 | H20→L | 88% |
| | | | H1→L2 | 8% | S34 | 3.99 | 0.00 | H5→L1 | 86% |
| | | | H1→L3 | 80% | S35 | 4.01 | 0.01 | H3→L1 | 17% |
| S19 | 3.29 | 0.00 | H12→L | 89% | | | | H1→L7 | 60% |
| | | | H1→L3 | 8% | | | | H→L7 | 9% |
| S20 | 3.33 | 0.00 | H13→L | 98% | S36 | 4.01 | 0.00 | H3→L1 | 81% |
| S21 | 3.43 | 0.00 | H15→L | 92% | | | | H1→L7 | 12% |
| S22 | 3.51 | 0.24 | H14→L | 69% | S37 | 4.02 | 0.00 | H17→L | 82% |
| | | | H→L4 | 26% | | | | H16→L | 9% |
| S23 | 3.55 | 0.00 | H18→L | 84% | S38 | 4.06 | 0.00 | H4→L1 | 94% |
| S24 | 3.55 | 0.18 | H14→L | 22% | S39 | 4.08 | 0.00 | H21→L | 91% |
| | | | H→L4 | 59% | S40 | 4.12 | 0.19 | H1→L8 | 36% |
| | | | | | | | | H→L8 | 46% |
| | | | | | | | | H→L9 | 8% |